\documentstyle[12pt,elsart12,epsf]{article}
\title{ Artificial Neural Network Methods
\\ in Quantum Mechanics}

\author{I. E. Lagaris, A. Likas and D. I. Fotiadis \\
Department of Computer Science \\
University of Ioannina \\
P.O. Box 1186 - GR 45110 Ioannina, Greece \\
Tel/fax: +30-651-48131, e-mail:lagaris@cs.uoi.gr\\
Correspondence: I.E. Lagaris}

\date{}

\begin{document}

\baselineskip 18pt

\maketitle

\vspace{0.25 cm}

\begin{center}
{\bf Abstract}
\end{center}

\noindent
In a previous article \cite{Lag96} we have shown how one  can employ Artificial
Neural Networks (ANNs) in order to solve non-homogeneous
ordinary and partial differential equations.
In the present work we consider the solution of eigenvalue problems
for differential and integrodifferential
operators, using ANNs.
We start by considering the Schr\"odinger equation for the Morse
potential that has an analytically known solution, to test the accuracy
of the method. We then proceed with
the Schr\"odinger and the Dirac equations for a muonic atom,
as well as with a non-local Schr\"odinger integrodifferential equation that
models the $n+\alpha$ system in the framework of the resonating group
method. 
In two dimensions we consider the well studied \cite{Sof} Henon-Heiles 
Hamiltonian and in three dimensions the model problem of
three coupled anharmonic oscillators.
The method in all of the treated cases proved to be highly accurate,
robust and efficient. Hence it is a promising tool for tackling
problems of higher complexity and dimensionality. 

\noindent {\bf PACS'96 codes}: 02.60.Lj, 02.60.Nm, 02.70.Jn, 03.65.Ge  

\newpage

\section{Introduction}

In a previous work \cite{Lag96} 
a general method has been presented for solving both 
ordinary differential equations (ODEs) and partial differential equations
(PDEs). This method 
relies on the function approximation 
capabilities of feedforward neural networks and leads to 
the construction of a solution written in a differentiable, closed analytic
form. The trial solution is suitably written so as to satisfy the
appropriate initial/boundary conditions and employs  
a feedforward neural network as the main 
approximation element. The  parameters of the network
(weights and biases) are then  adjusted so as
to minimize a suitable  error function, which in turn 
is equivalent to satisfying the differential equation
at selected points in the definition domain.

There are many results both theoretical and experimental that
testify for the approximation capabililities of neural networks 
\cite{Fun89,Hor89,Wil95}.
The most important one is that a
feedforward neural network with one hidden layer 
can approximate any function to arbitrary accuracy 
by appropriately increasing the number of units in
the hidden layer \cite{Hor89}. This fact has led us 
to consider this type of network architecture  
as a candidate model for treating differential equations. 
In fact the employment of neural networks 
as a tool for solving differential equations
has many attractive features \cite{Lag96}:
\begin{itemize}
\item The solution via neural networks
is a {\em differentiable, closed analytic form} 
easily used in any subsequent calculation with superior 
interpolation capabilities. 
\item Compact solution models are obtained due to the small number
of required parameters. This fact also results in low memory demands.
\item There is the possibility of direct hardware implementation
of the method on specialized VLSI chips called {\em neuroprocessors}. 
In such a case there will be a tremendous increase 
in the processing speed that  will offer the opportunity 
to tackle many difficult high-dimensional problems requiring 
a large number of grid points.  Alternatively,
it is also possible for the proposed method to be efficiently 
implemented on parallel architectures.  
\end{itemize}

In this paper we present a novel technique for solving 
eigenvalue problems of differential and integrodifferential 
operators, in one, two and three dimensions, 
that is based on the use of MLPs for the parametrization 
of the solution, on the collocation method for the formulation
of the error function and on optimization procedures. 

All the problems we tackle come from the field of Quantum Mechanics, 
i.e. we solve mainly Schr\"odinger problems and we have applied the 
same technique to the Dirac equation that is reduced to a system of 
coupled ODEs.
In addition, for the Schr\"odinger equation one can employ
the Raleigh-Ritz variational principle, where again the variational
trial wavefunction is parametrized using MLPs.
For the two-dimensional Hennon-Heiles potential, we compare the 
resulting variational and the collocation solutions.

A description of the general formulation of the proposed
approach is presented in section 2.
Section 3 illustrates several cases of 
problems where the proposed technique has been applied along with 
details concerning the implementation of the method
and the accuracy of the obtained solution. In addition, in a two dimensional
problem, we provide
a comparison of our results with those obtained by 
a solution  based on finite elements. 
Finally, section 4 contains conclusions and directions for future research.

\section{The Method}

Consider the following differential equation: 
\begin{equation}
H \Psi(\vec{r})=f(\vec{r}),\ \ \  in \ \ D
\end{equation} 
\begin{equation}
\Psi(\vec{r})=0,\ \ \  on \ \ \partial D
\end{equation}
where  $H$ is a linear differential operator, $f(\vec{r})$ is a known
function, $D \subset R^3$ 
and $\partial D$ is the boundary of $D$. Moreover, we denote
$\bar{D} = D \cup \partial D$. We assume that $f \in C(\bar{D})$ and      
the solution
$\Psi(\vec{r})$ belongs to $C^k(\bar{D})$, the space of continuous
functions with continuous partial derivatives up to $k$ order inclusive
($k$ is the higher order derivative appearing in the operator $H$,
$H \Psi(\vec{r}) \in C(\bar{D}))$.  The set of the admissible functions
$$
\{ \Psi(\vec{r}) \in C^k(\bar{D}), \ \ \vec{r} \in D \subset R^3, 
\ \ \Psi(\vec{r})=0 \ \ on \ \ \partial D \}  
$$
forms a linear space. In the present analysis we also assume that the
domain under consideration $D$ is bounded and its boundary $\partial D$
is sufficiently smooth (Lipschitzian).

In order to solve this problem we have
proposed a technique \cite{Lag96}, that considers a trial solution of
the form $\Psi_t (\vec{r})=A(\vec{r})+
B(\vec{r},\vec{\lambda})N(\vec{r},\vec{p})$ 
which employs a feedforward neural network with parameter vector
$\vec{p}$ (to be adjusted). The parameter vector $\vec{\lambda}$
should also be adjusted during minimization. The specification 
of functions $A$ and $B$ should be done so that $\Psi_t$ satisfies 
the boundary conditions 
regardless of the values of
$\vec{p}$ and $\vec{\lambda}$. 
 
To obtain a solution to the above differential equation
the collocation method has been employed \cite{Kin91}
which assumes a discretization of the domain $D$ 
into a set points $\vec{r_i}$. The problem is then
transformed into a minimization one with respect to the 
parameter vectors $\vec{p}$ and $\vec{\lambda}$:
\begin{equation}
min_{\vec{p},\vec{\lambda}}\sum_i 
[H\Psi_t(\vec{r_i}) -f(\vec{r_i})]^2
\end{equation}
If the obtained minimum has a value close to zero, then we consider that an
approximate solution has been recovered. 

Consider now the case of the following general eigenvalue
problem:
\begin{equation}
H \Psi(\vec{r})=\epsilon \Psi(\vec{r}),\ \ \  in \ \ D
\end{equation}
\begin{equation}
\Psi(\vec{r})=0,\ \ \  on \ \ \partial D
\end{equation}
In this case a trial solution may take the form:
$\Psi_t(\vec{r})=B(\vec{r},\vec{\lambda})
N(\vec{r},\vec{p})$ where $B(\vec{r},\vec{\lambda})$
is zero on $\partial D$, for a range of values of $\vec{\lambda}$.
By discretizing the domain, the problem is transformed
to minimizing the following error quantity, with respect
to the parameters $\vec{p}$ and $\vec{\lambda}$: 
\begin{equation}
Error(\vec{p},\vec{\lambda})=
\frac{
\sum_i [H\Psi_t(\vec{r_i},\vec{p},\vec{\lambda})-
\epsilon \Psi_t(\vec{r_i},\vec{p},\vec{\lambda})]^2
}{ 
\int |\Psi_t|^2 d\vec{r}}
\end{equation}
where $\epsilon$ is computed as:
\begin{equation}
\epsilon=\frac{\int \Psi_t^{\star}H\Psi_t d\vec{r}}
{\int |\Psi_t|^2 d\vec{r}}
\end{equation}

A method similar in spirit has been proposed long ago by Frost et al \cite{Frost}
and is known as the "Local Energy Method".
In the proposed approach the trial solution $\Psi_t$ employs 
a feedforward neural network and more specifically a multilayer
perceptron (MLP).
The parameter vector $\vec{p}$ corresponds to the
weights and biases of the neural architecture. Although it is
possible for the MLP to have many hidden layers we have considered
here the simple case of single hidden layer MLPs, which have been proved
adequate for our test problems.  

Consider a multilayer perceptron with $n$ input units, 
one hidden layer with $m$ sigmoid units and a linear output unit
(Figure 1). 
The  extension to the case of more than one hidden layers can be 
obtained accordingly.
For a given input vector $\vec{r}=(r_1,
\ldots, r_n)$ the output of the network is $N=\sum_{i=1}^m v_i
\sigma(z_i)$ where $z_i= \sum_{j=1}^n w_{ij}r_j+u_i$, $w_{ij}$ denotes 
the weight from the input unit $j$ to the
hidden unit $i$, 
$v_i$ the weight from the hidden unit $i$ to the output,
$u_i$ the bias of hidden unit $i$ and $\sigma(z)$ the 
sigmoid transfer function: $\sigma(z)=1/(1+\exp(-z))$.
It is straightforward to show that \cite{Lag96} :
\begin{equation}
\frac{\partial^k N}{\partial r^{k}_j}=\sum_{i=1}^{m} v_iw_{ij}^k \sigma_i^{(k)}
\end{equation}
where $\sigma_i=\sigma(z_i)$ and $\sigma^{(k)}$ denotes
the $k^{th}$ order derivative of the sigmoid.
Moreover it is readily verifiable that:
\begin{equation}
\frac{\partial^{\lambda_1}}{\partial r_1^{\lambda_1}}
\frac{\partial^{\lambda_2}}{\partial r_2^{\lambda_2}}
\ldots
\frac{\partial^{\lambda_n}}{\partial r_2^{\lambda_n}}N=\sum_{i=1}^m v_iP_i
\sigma_i^{(\Lambda)}
\end{equation}
where 
\begin{equation}
P_i=\prod_{k=1}^n w_{ik}^{\lambda_k}
\end{equation}
and $\Lambda=\sum_{i=1}^n \lambda_i$.

Once the derivative of the error with respect
to the network parameters has been defined
it is then straightforward to employ almost any minimization technique.
For example it is possible to use either the steepest  
descent (i.e. the backpropagation algorithm or any of its variants), or 
the conjugate gradient method or other techniques proposed in the literature.
We used the MERLIN optimization package \cite{Mer1,Mer2} for our experiments,
where many algorithms are available. We mention in passing
that the BFGS method has demonstrated outstanding performance.
Note that for a given grid point the calculation of the gradient of 
each network with respect to the adjustable parameters,  
lends itself to parallel computation.

Using the above approach it is possible to
calculate any number of states. 
This is done by projecting out from the
trial wavefunction the already  computed levels. 

If $| \Psi_0 >, | \Psi_1 >, \dots, | \Psi_k > $ are
computed orthonormal states, a trial state $| \Psi_t >$ 
orthogonal to all of them can be obtained by 
projecting out their components 
from a general function $| \tilde{\Psi_t}> $
that respects the boundary conditions, namely:
$$
| \Psi_t > = (1 - | \Psi_0 > < \Psi_0 |) (1 - | \Psi_1 > < \Psi_1 |)
\dots (1 - | \Psi_k > < \Psi_k |) | \tilde{\Psi_t} >
$$
$$
=  (1 - | \Psi_0 > < \Psi_0 | - | \Psi_1 > < \Psi_1 |
\dots - | \Psi_k > < \Psi_k |) | \tilde{\Psi_t} >
$$

\section{Examples}

\subsection{Schr\"odinger equation for the Morse Potential} 
The Morse Hamiltonian for the $I_2$ -- molecule in the 
atomic units system, is given by:
$$  H = -\frac{1}{2 \mu}\frac{d^2}{dx^2} + V(x) $$
where $V(x) = D[e^{-2\alpha x} -2 e^{-\alpha x} + 1 ] $
and $ D= 0.0224, \ \ \alpha = 0.9374, \ \ \mu = 119406$.

\noindent
The energy levels are known analytically \cite{Flu}, and are
given by:\\
\noindent $\epsilon_n = (n + \frac{1}{2})(1-\frac{n+ 1/2}{\zeta})\xi $
with $\zeta = 156.047612535,\ \ \xi = 5.741837286\  10^{-4} $.
The ground state energy is $ \epsilon_0 = 0.286171979 \ 10^{-3}$.
We parametrize as:
$$
\phi_t (x) =  e^{-\beta x^2} N (x, \vec{u}, \vec{w}, \vec{v}),\ \  \beta >0
$$
with $N$ being a feedforward artificial neural network with one
hidden layer and $m$ sigmoid hidden units, ie:
\[
N (x, \vec{u}, \vec{w}, \vec{v}) = \sum_{j=1}^{m} v_j \sigma (w_j x +u_j)
\]
We solve the problem in the interval  $ \  -1 \le r \le 2 $ using
150 equidistant grid points with $m=8$.
We minimize the quantity:
$$ \frac{1}{\int \phi_t^2 (x) dx } 
\sum_i [H \phi_t (x_i)- \epsilon\phi_t (x_i)]^2 $$
where $\epsilon = \frac{\int \phi_t (x) 
H \phi_t (x) dx }{\int \phi_t^2 (x) dx }$.
We find for the ground state energy the value  
$0.286171981 \ 10^{-3}$ which is
in excelent agreement with the exact
analytical result.

\subsection{Schr\"odinger equation for muonic atoms} 

The $s$-state equation for the reduced radial wavefunction $\phi (r) = r R(r)$
of a muon in the field of a nucleus is:
$$
- \frac{\hbar^2}{2\mu} \frac{d^2 \phi}{d r^2} (r) + V (r) \phi (r) = \epsilon \phi (r)
$$
\noindent
with:  $\phi(r=0) = 0$ and  $\phi (r) \sim e^{-kr}$, $k > 0$ 
for a bound state. \\
\noindent
$\mu$ is the reduced muon mass given by:
$
\frac{1}{\mu} = \frac{1}{m_{\mu}} + \frac{1}{Z m_p + N m_n}
$, \ \ 
where $m_{\mu}$ is the muon mass and $m_p, m_n$ the masses of the
proton and neutron respectively. 
$Z$ is the number of protons and N the number of neutrons for the nucleus
under consideration.
(In our example we calculate the muonic wavefunction in $_{82}Pb^{208}$).

\noindent
The potential  has two parts, i.e.:
$ V(r) = V_e(r) + V_p(r) $, \ \ where
$$V_e (r) = -e^2 \int \frac{\rho(r')}{\| \vec{r} - \vec{r'} \|} d^3 r'$$
is the electrostatic potential,
$\rho(r)$ is the proton number-density given by  
$$
\rho(r)= A/(1 + e^{(r-b)/c}) $$
with $A=0.0614932,\ \  b=6.685$ and \  $ c= 0.545 $ \ \  and 
$$
V_p(r) = \frac{2\alpha}{3\pi} \left[ V_L (r) -\frac{5}{6} V_e (r) \right] 
$$
is the effective potential due to vacuum polarization \cite{Vacuum}
with $\alpha= \frac{1}{137.037}$ the fine-structure constant. 
\[
V_L(r) = -2\pi \frac{e^2}{r} \int_0^\infty \rho(r') r' \left\{ |r - r'| 
 [ln (C |r - r'| \lambda_e) -1] - (r+r') [ln (C (r+r')/\lambda_e -1] \right\} dr' 
\]

\noindent
with $C=1.781$ and $\lambda_e$ the electron Compton wavelength
divided by $2\pi$.

\noindent
We parametrized the trial wavefunction as:
$$
\phi_t (r) = r e^{-\beta r} N (r, \vec{u}, \vec{w}, \vec{v}),\ \  \beta >0
$$
where again $N$ is again a feedforward artificial neural network with one
hidden layer having 8 sigmoid hidden units.

\noindent
The energy eigenvalue is calculated as:
\[
\epsilon = 
\frac{1}{\int_{0}^{\infty} \phi_t^2 (r) dr}
\left[ \frac{\hbar^2}{2\mu} \int_{0}^{\infty} (\frac{d\phi_t}{dr})^2 
dr + \int_{0}^{\infty} V(r) \phi_t^2 (r) dr \right]  
\]
The intergrals have been calculated using the Gauss-Legendre rule.
We used 80 points in the range $[0,40]$.
The quantity:
\[
\frac{1}{\int_{0}^{\infty} \phi_t^2 (r) dr}
\sum_{i} \left\{ -\frac{\hbar^2}{2\mu} \frac{d^2}{dr^2} \phi_t (r_i)
+V(r_i) \phi_t(r_i) -\epsilon \phi_t (r_i) \right\}^2 
\]
is being minimized with respect to $\vec{u}, \vec{w}, \vec{v}$.

\noindent
We used for ${r_i}$ the same points as in the Gauss-Legendre 
Integration. 
We obtained for the energy $\epsilon = -10.47 MeV$.
The radial wavefunction $\frac{1}{r}\phi(r)$ is shown in Fig. 2c.
%
%

\subsection{Dirac equation for muonic atoms}

The relativistic Dirac
$s$-state equations for the small and large parts of the reduced radial
wavefunction of a muon bound by a nucleus are \cite{Dirac}: 
$$
\frac{d}{dr}f(r) + \frac{1}{r}f(r) = \frac{1}{\hbar c}(\mu c^2 -E+V(r))g(r)
$$
$$
\frac{d}{dr}g(r) - \frac{1}{r}g(r) = \frac{1}{\hbar c}(\mu c^2 +E-V(r))f(r)
$$
\noindent
with $\mu$ and $ V(r)$ being as in the previous example.

The total energy $E$ is calculated by:
$$
E = \frac{1}{\int_0^\infty [g^2(r)-f^2(r)]dr}
\{ \mu c^2\int_0^\infty [g^2(r)+f^2(r)]dr +
 \int_0^\infty V(r)[g^2(r)-f^2(r)]dr\}
$$


We parametrized the trial solutions $f_t(r)$  and $g_t(r)$ as:
$$
f_t(r) = r e^{-\beta r} N (r, \vec{u_f}, \vec{w_f}, \vec{v_f}),\ \  \beta >0
$$
$$
g_t(r) = r e^{-\beta r} N (r, \vec{u_g}, \vec{w_g}, \vec{v_g}),\ \  \beta >0
$$
and minimized the following error quantity:
\[
\frac{
\sum_i\{ [\frac{d f(r_i)}{dr} + \frac{f(r_i)}{r_i}
- \frac{\mu c^2 -E+V(r_i)}{\hbar c}g(r_i)]^2  +
[\frac{d g(r_i)}{dr} - \frac{g(r_i)}{r_i}
- \frac{\mu c^2 +E-V(r_i)}{\hbar c}f(r_i)]^2 \}
}
{\int_0^\infty [g^2(r)+f^2(r)]dr}
\]
The binding energy is given by $\epsilon = E - \mu c^2$.
We find $\epsilon = -10.536 \ MeV$.
The small and the large parts of the radial wavefunction
$\frac{1}{r}f(r)$ and $\frac{1}{r}g(r)$  are  shown in Fig. 2a
and 2b,
along with the Schr\"odinger radial wavefunction (Fig. 2c). 
The integrals and the training were performed using the same 
points as in the previous example.

\subsection{Non-Local Schr\"odinger equation for the $n+\alpha$ system} 

We consider here the non-local Schr\"odinger equation :
$$
- \frac{\hbar^2}{2\mu} \frac{d^2 \phi}{d r^2} (r) + V (r) \phi (r) 
+ \int_0^\infty K_0(r,r') \phi (r') dr'
= \epsilon \phi (r)
$$
with $V(r)= -V_0 e^{-\beta r^2}$,  where
$V_0=41.28386$, $ \beta =  0.2751965$
and \\
\noindent $K_0(r,r')= -A e^{-\gamma (r^2+r'^2)}(e^{2 k r r'}-e^{-2 k r r'})$
with $ A =    -62.03772$ , $\gamma =   -0.8025$, 
$ k =    0.46$.
This describes the $n+\alpha $ system and is derived in the framework 
of the Resonating Group Method \cite{Tang}, $\mu$ is the system's 
reduced mass given by:
$\frac{1}{\mu}=\frac{1}{m_n} + \frac{1}{2 m_n + 2 m_p}$.


We parametrized the trial wavefunction as:
$$
\phi_t (r) = r e^{-\beta r} N (r, \vec{u}, \vec{w}, \vec{v}),\ \ \beta >0
$$
where the neural architecture is the same as in the previous cases
and minimized the following error quantity:
\[
\frac{
\sum_i \left\{ - \frac{\hbar^2}{2 \mu} \frac{d^2}{d r^2} 
\phi_t (r_i) + V(r_i) \phi_t (r_i) + \int_0^\infty K_0 (r_i, r')  \phi_t
(r') dr' - \epsilon \phi_t (r_i) \right\}^2
}
{\int_0^\infty  \phi^2_t (r) dr}
\]
where the energy is estimated by:
\[
\epsilon = \frac{\hbar^2/2\mu \int_0^\infty  (\frac{d\phi_t}{dr})^2 dr + 
\int_0^\infty  V(r) \phi^2_t (r) dr +
 \int_0^\infty  \int_0^\infty  K_o (r,r') \phi_t (r) \phi_t (r') dr dr'}
{\int_0^\infty  \phi^2_t (r) dr}
\]
We have considered 100 equidistant points in $[0,12]$ and the
computed ground state is depicted in Fig. 3, while the corresponding
eigenvalue was found equal to -24.07644, in agreement with 
previous calculations \cite{Sof}.

\subsection{Two dimensional Schr\"odinger equation}

We consider here the well studied \cite{Sof}  example of the Henon-Heiles 
potential.

\noindent
The Hamiltonian is written as:
$$
H= - \frac{1}{2} (\frac{\partial^2}{\partial x^2} +
\frac{\partial^2}{\partial y^2}) + V(x,y)
$$
with $V(x,y) = \frac{1}{2} (x^2 + y^2) + \frac{1}{4 \sqrt{5}}
(xy^2 - \frac{1}{3} x^3) $.


\noindent
We parametrize the trial solution as:
$$
\phi_t (x,y) = e^{-\lambda (x^2 + y^2)} N (x, y, \vec{u}, \vec{w}^{(x)}, 
\vec{w}^{(y)}, \vec{v}), \quad \lambda > 0
$$
where N is a feedforward neural network with one hidden layer (with
$m=8$ sigmoid hidden units) and
two input nodes (accepting the x and y values).
\[
N (x, y, \vec{u}, \vec{w}^{(x)}, \vec{w}^{(y)}, \vec{v}) = 
\sum_{j=1}^{m} v_j \sigma (x w_j^{(x)} + y w^{(y)}_j + u_j)
\]
We have considered a grid of $20 \times 20 $ points in $[-6,6]\times [-6,6]$. 
The quantity minimized is:
\begin{equation}
\sum_{i,j} [ H \phi_t (x_i, y_j) - \epsilon \phi_t (x_i, y_j) ]^2 /
\int_{-\infty}^{\infty} \int_{-\infty}^{\infty} dx dy \phi^2_t (x, y)
\end{equation}
where the energy is calculated by:
\[
\epsilon = \frac{\int_{-\infty}^{\infty} \int_{-\infty}^{\infty}
\phi_t (x,y) H \phi_t (x,y) dx dy}{\int_{-\infty}^{\infty} 
\int_{-\infty}^{\infty} \phi_t^2 (x,y)  dx dy}
\]




For this problem we calculate not only the ground state but a
few more levels. The way we followed is the extraction from the
trial wavefunction of the already  computed levels as 
described in Section 2. If for example
by $\phi_0 (x,y)$ we denote the normalized ground state, the
trial wavefunction to be used for the computation of another level 
would be:
\[
\phi_t (x,y) = \tilde{\phi}_t (x,y) - \phi_0 (x,y)
 \int_{-\infty}^{\infty}  \int_{-\infty}^{\infty} 
 \phi_0 (x',y') \tilde{\phi}_t (x',y') dx' dy' 
\]
where $\tilde{\phi}_t (x,y)$ is parametrized in the same way as
before.

Note that $\phi_t (x,y)$ is orthogonal to $\phi_0 (x,y)$ by
construction. Following this procedure we 
calculated the first four levels for
the Henon-Heiles Hamiltonian.
Our results are reported in Figs. 4-7.



We also calculated the variational ground state wave-function 
for this problem by minimizing the expectation value of the
Hamiltonian, using an identical neural form.
In Figs. 8--9, we plot the pointwise error, i.e. the (normalized)
summand of eq. (11) for  
the collocation and the variational wavefunctions respectively.

\subsection{Three Coupled Anharmonic Oscillators}
As a three-dimensional example we consider the potential
for the three coupled sextic anharmonic oscillators \cite{Kal}:
\[
V(x,y,z)=V(x)+V(y)+V(z)+xy+xz+yz
\]
where 
\[
V(x)=\frac{1}{2}x^2 + 2x^4 + \frac{1}{2}x^6
\]

The trial solution $\phi_t(x,y,z)$ is parametrized as: 
$$
\phi_t (x,y,z) = e^{-\lambda (x^2 + y^2 + z^2)} 
N (x, y, z, \vec{u}, \vec{w}^{(x)},
\vec{w}^{(y)}, \vec{w}^{(z)}, \vec{v}), \quad \lambda > 0
$$
where N is a feedforward neural network with one hidden layer 
(with $m=25$ hidden units) and
three input nodes (accepting the values of x, y and z):
\[
N (x, y, z, \vec{u}, \vec{w}^{(x)}, \vec{w}^{(y)}, 
\vec{w}^{(z)}, \vec{v}) =
\sum_{j=1}^{m} v_j \sigma (x w_j^{(x)} + y w^{(y)}_j + z w^{(z)}_j + u_j)
\]
We have considered a 28$\times$28$\times$28 grid 
in the $[-4,4]\times [-4,4]\times [-4,4]$ 
domain both for computing the integrals and calculating 
the following error quantity
that was minimized:
\[
\sum_{i,j,k} [ H \phi_t (x_i, y_j, z_k) - \epsilon 
\phi_t (x_i, y_j, z_k) ]^2 /
\int_{-\infty}^{\infty} \int_{-\infty}^{\infty} 
\int_{-\infty}^{\infty} \phi^2_t (x, y, z) dx dy dz
\]
where the energy is calculated by:
\[
\epsilon = \frac{\int_{-\infty}^{\infty} \int_{-\infty}^{\infty} 
\int_{-\infty}^{\infty}
\phi_t (x,y,z) H \phi_t (x,y,z) dx dy dz}{\int_{-\infty}^{\infty}
\int_{-\infty}^{\infty} \int_{-\infty}^{\infty}
\phi_t^2 (x,y,z)  dx dy dz}
\]
The ground state was computed and the corresponding eigenvalue was
found equal to 2.9783, in agreement with the highly accurate
result obtained by Kaluza \cite{Kal}.

\section{Finite Element Approach}

The two-dimensional Schr\"ondiger equation for the Henon-Heiles
potential was also solved using the finite element approach
in which the solution is expressed in terms of piecewise continuous
biquadratic basis functions:
\begin{equation}
\psi=\sum_{i=1}\psi_i \Phi_i(\xi,n)
\end{equation}
where $\Phi_i$ is the biquadratic basis function and $\psi_i$ is
the unknown at the $i$-th node of the element. The physical
domain $(x,y)$ is mapped on the computational domain $(\xi,n)$
through the isoparametric mapping:
\begin{equation}
x=\sum_{i=1} x_i \Phi_i(\xi,n)
\end{equation}
\begin{equation}
y=\sum_{i=1} y_i \Phi_i(\xi,n)
\end{equation}
where $\xi$ and $n$ are the local coordinates in the computational
domain $(0 \leq \xi, n \leq 1)$ and $x_i$, $y_i$ the $i$-th node
coordinates in the physical domain for the mapped element.

The Galerkin Finite Element formulation calls for the weighted
residuals $R_i$ to vanish at each nodal point $i=1, \ldots, N$ :
\begin{equation}
R_i=\int_{\Omega} (H\psi-e\psi)\Phi_i det({\bf J}) d\xi dn=0 
\end{equation}
where
${\bf J}$ the Jacobian of the isoparametric mapping with
\begin{equation}
det({\bf J})=\frac{\partial x}{\partial \xi} \frac{\partial y}{\partial n}-
       \frac{\partial x}{\partial n} \frac{\partial y}{\partial \xi}
\end{equation}
These requirements along with the imposed boundary conditions 
constitute a system of linear equations which can be written in
a matrix form as:
\begin{equation}
{\bf K \psi} = \epsilon {\bf M \psi}
\end{equation}
where ${\bf K}$ is the stiffness and  ${\bf M}$ is the mass matrix.
The stiffness matrix in its
local element form is:
\begin{equation}
\int \int \{ \frac{1}{2} [\frac{\partial\Phi_i}{\partial x} \frac{\partial\Phi_j}{\partial x}
+ \frac{\partial\Phi_i}{\partial y} \frac{\partial\Phi_j}{\partial y}] +
\frac{1}{2}(x^2+y^2)\Phi_i\Phi_j  + \frac{1}{4\sqrt{5}}(xy^2-\frac{1}{3}x^3)\Phi_i\Phi_j \}
det({\bf J}) d\xi d\eta
\end{equation}
The matrix ${\bf M}$ obtained above in its local element form is:
\begin{equation}
\int_{\Omega} \Phi_i \Phi_j det({\bf J})d\xi dn
\end{equation}  
Due to the Dirichlet boundary conditions zeros appear in the
diagonal. Thus the mass matrix is singular and the total number of
zeros in the diagonal of the global matrix is equal to the number of
nodes on the boundaries and its degree of singularity depends on the
size of the mesh. 

\subsection{Extracting Eigenvalues and Eigenvectors}

For the problem under discussion only the eigenvalues of the
generalized eigenvalue problem with the smallest real parts are needed.
The eigenvalue problem is a symmetric one generalized eigenvalue
problem but for generality purposes it is solved as a nonsymmetric
one. Due to the size of the problem (from 1000 - 4000 unknowns in
our solution) direct methods are not suitable.

We use Arnoldi's method as it has been implemented by Saad 
\cite{Saad1,Saad2,Saad3}, which is
based on an iterative deflated Arnoldi's algorithm. 
Saad proposes an iterative improvement of the eigenvectors
as well as a Schur-Wiedland deflation to overcome
cancellation errors in the orthonormalization of the eigenvectors
at each step due to the finite arithmetic.

\begin{table}
\begin{center}
\begin{tabular}{||c|c|c|c|c|c||} \cline{1-6}
     \multicolumn{1}{||c|}{5$\times$5}
   & \multicolumn{1}{c|}{7$\times$7}
   & \multicolumn{1}{c|}{11$\times$11}
   & \multicolumn{1}{c|}{16$\times$16}
   & \multicolumn{1}{c|}{21$\times$21}
   & \multicolumn{1}{c||}{29$\times$29} \\ \hline
  1.0075  & 0.9997  &  1.0015  &  0.9994  & 0.9989 & 0.9986\\ \hline
  2.1988  & 2.0852 &   2.0037  &  1.9930  & 1.9911 & 1.9901\\ \hline
  2.2001  & 2.0862  &  2.0037  &  1.9930  & 1.9911 & 1.9901\\ \hline
  3.2495  & 3.0159  &  2.9767  &  2.9648  & 2.9593 & 2.9571\\ \hline
  3.2878  & 3.0515  &  3.0065  &  2.9943  & 2.9885 & 2.9857\\ \hline
  4.4347  & 4.1139  &  3.9868  &  3.9433  & 3.9323 & 3.9262\\ \hline
\end{tabular}
\end{center}
\caption{Computed eigenvalues of the Henon-Heiles Hamiltonian
using the FEM approach for various mesh sizes.}
\end{table}

If ${\bf K}$ is nonsingular, a simple way to handle the generalized
eigenvalue problem is to consider the "reciprocal" problem:
\begin{equation}
{\bf M\psi}=\mu {\bf K\psi}
\end{equation}  
where $\mu=1/\epsilon$. The infinite valued eigenvalues are transformed
into zero eigenvalues. However, 
due to computer round-off errors, the infinite-valued
eigenvalues actually correspond to very large values in the
calculations, which are turned into very small valued eigenvalues and 
not to exact zeros in the reciprocal problem.



An alternative method would require the elimination of the rows with
zero diagonal in the mass matrix, which are the rows corresponding
essentialy to the boundary conditions. This
scheme requires a number of manipulative operations
on ${\bf K}$ and ${\bf M}$ which are prohibitive for large systems.
The method is called the `reduced algorithm' and requires the storage
of the stiffness and the mass matrix. Other techniques
have been proposed and mainly are transformations of the 
generalized eigenvalue problem that map the infinite eigenvalues to
one or more specified points in the complex plane \cite{Chris}. 
The Shift-and-Invert
transformation maps the infinite eigenvalues to zero.
In the problem under discussion we have used the transformation:
\begin{equation} 
{\bf C}=({\bf M}-\sigma{\bf K})^{-1}{\bf K}
\end{equation}
and the problem (20) is transformed to the problem:
\begin{equation}
{\bf C \psi}=\mu'{\bf \psi}
\end{equation}
whose eigenvalues are related to those of equation (20) through
the relation $\mu'=1/(\mu-\sigma)$, where $\sigma$ is a real number
called shift. This transformation favors eigenvalues with real part
close to the shift. The eigenvalues $\epsilon$ of the original problem
are then given by $\epsilon=\mu'/(1+\sigma \mu')$.

The generalized eigenvalue problem was solved on a rectangular domain.
Figs. 10 and 11 shows the evolution of the first and second 
eigenvalues as
the number of equidistributed elements of the mesh and consequently the
number of unknowns increases.
Convergence occurs for a grid of equal elements $ ( 29 \times 29 )$
which results in a system of $3481$ unknowns.
The convergence of the first six eigenvalues is also shown in Table 1. 
It is obvious that in order to get accurate eigenvalues, dense FEM meshes
must be used and this limits the application of the method.

\section{Conclusions}

We presented a novel method appropriate for solving 
eigenvalue problems of ordinary, partial 
and integrodifferential equations.
We checked the accuracy of the method by comparing to
a result that is analyticaly known, i.e. the ground
state energy of the Morse Hamiltonian.
We then applied the method to two realistic and interesting
problems, namely to the {\em Schr\"odinger } and to the {\em Dirac }
equations for a muonic atom. In these equations we take account
of the finite protonic charge distribution as well as of the 
{\em Vacuum Polarization } effective potential.
Preliminary calculations using a proton density delivered by
Quasi-RPA have also been performed \cite{Kosmas}.
Since both the {\em Schr\"odinger } and the {\em Dirac } equations
can be solved analyticaly in the case of a point charge nucleus,
(ignoring also the vacuum polarization correction),
we conducted  calculations (not reported in this article)
and determined the energies  
for the $4f$ and $5g$ levels to within $1 \ ppm$ \cite{Anag}.
The wide applicability of the method is shown by solving 
an integrodifferential problem, coming from the field
of Nuclear Physics.
The two dimensional benchmark, namely the Henon--Heiles 
Hamiltonian, that has been considered by many authors and
solved by a host of methods, was considered as well. 
Here we obtained not only the ground state, but also some of the 
higher states, following a projection technique to supress
the already calculated levels. Our results are in excellent agreement
with the ones reported in the literature.
We solved this problem also by a standard finite element technique
and we compared the computational resources and effort. 
It is clear that the present method  is far more economical
and efficient. Also, as we have previously shown \cite{Lag96}
for the case of non-homogeneous equations, its interpolation
capabilities are superb.
Coming to an end, we solved a three-dimensional problem that
imposes a heavier load. Again the results for the three-coupled
anharmonic sextic oscillators are in agreement with the high precision 
ones obtained in \cite{Kal} by a semi-analytical method.
The examples treated in this article are essentially single particle
problems. (In example 3.4 the few-body nature is embeded in the
non-local kernel). Many-body problems will impose a much heavier
computational load, and hence the fast convergence property of 
the sigmoidal functions \cite{Baron}, as well as the availability
of specialized hardware become very important.
Few-body problems may be handled by extending the method in a rather
straightforward fashion. However for many-body problems it is not clear
as of yet how to find a tractable neural form for the trial wavefunction.
The method is new and of course there is room for further
research and development. Issues that will occupy us in the future
are optimal selection of the training set, networks with more than
one hidden layers, radial basis function networks few-body systems 
and implementation on specialized neural hardware.

We would like to acknowledge the anonymous referee for his useful 
suggestions that resulted in making the article more valuable.

\clearpage

\begin{figure}
\caption{Feedforward neural network with one
hidden layer.}
\end{figure}

\begin{figure}
\caption{Ground state of:
(a),(b) the Dirac and (c) the Schr\"odinger
equation  for muonic atoms.}
\end{figure}

\begin{figure}
\caption{Ground state of the non-local Schr\"odinger
equation for the $n+\alpha$ system ($\epsilon=-24.07644$).}
\end{figure}

\begin{figure}
\caption{Ground state of the Henon-Heiles problem
($\epsilon=0.99866$).}
\end{figure}

\begin{figure}
\caption{First excited state of the Henon-Heiles problem
($\epsilon=1.990107$).}
\end{figure}

\begin{figure}
\caption{Second excited state (degenerate) of the Henon-Heiles problem
($\epsilon=1.990107$).}
\end{figure}

\begin{figure}
\caption{Third excited state of the Henon-Heiles problem
($\epsilon=2.957225$).}
\end{figure}

\begin{figure}
\caption{Pointwise normalized error for the collocation
wavefunction.}
\end{figure}

\begin{figure}
\caption{Pointwise normalized error for the variational
wavefunction.}
\end{figure}

\begin{figure}
\caption{Convergence of the first
eigenvalue as a function
of the mesh size (number of unknowns).}
\end{figure}

\begin{figure}
\caption{Convergence of the second
eigenvalue as a function
of the mesh size (number of unknowns).}
\end{figure}

\clearpage

\begin{figure}
\centerline{\epsfysize=10cm\epsffile{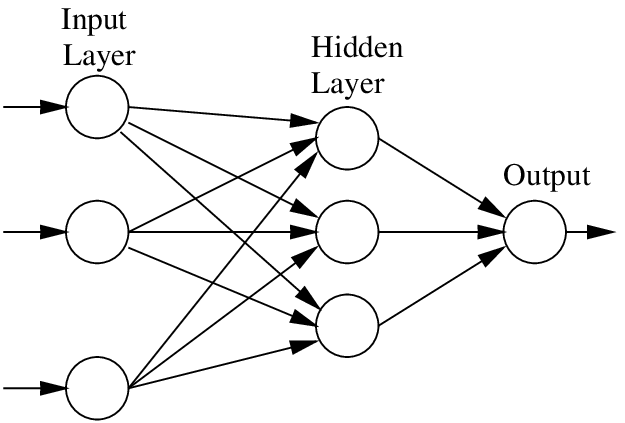}}
\end{figure}

\clearpage

\begin{figure}
\centerline{\epsfysize=10cm\epsffile{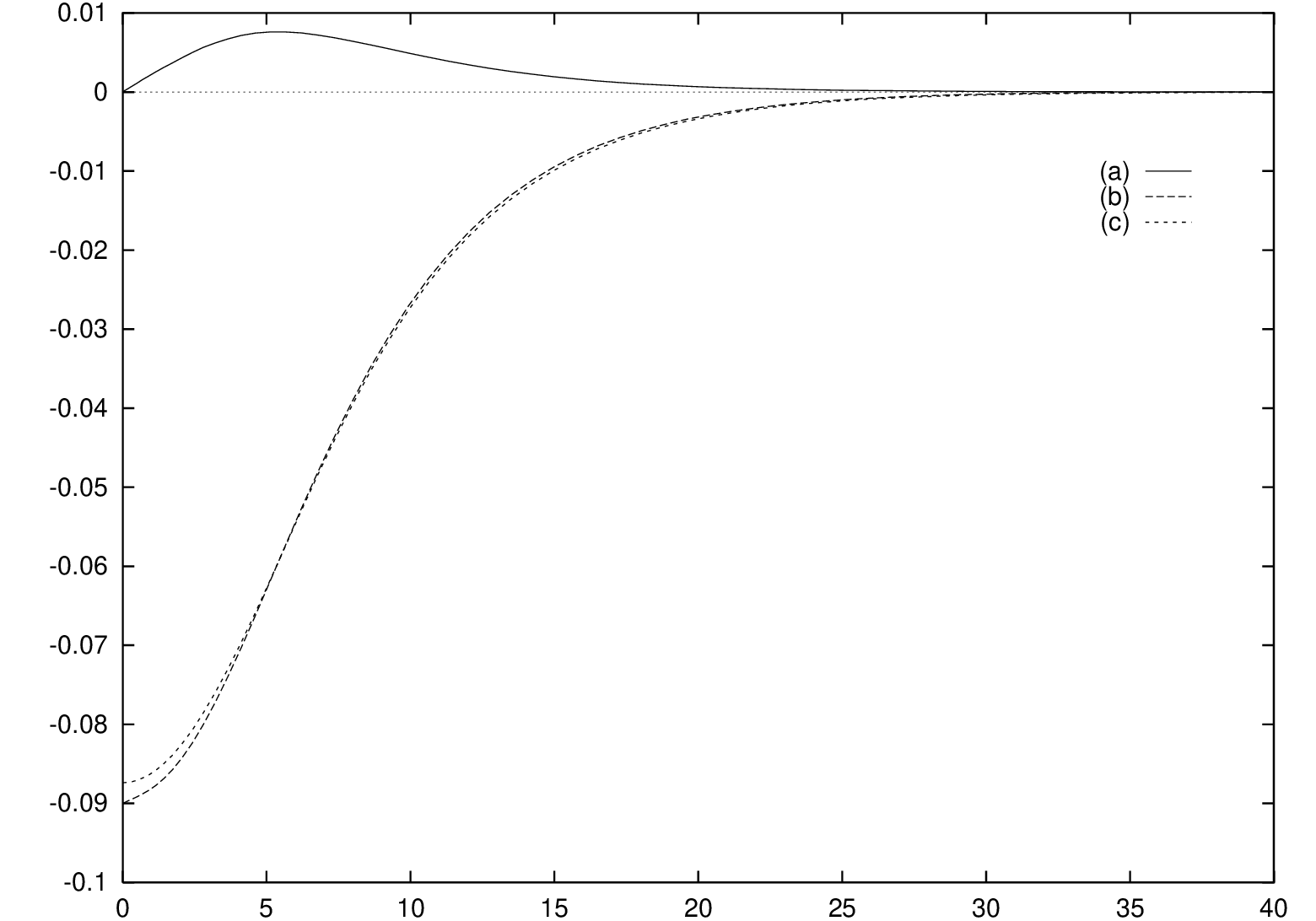}}
\end{figure}

\clearpage

\begin{figure}
\centerline{\epsfysize=10cm\epsffile{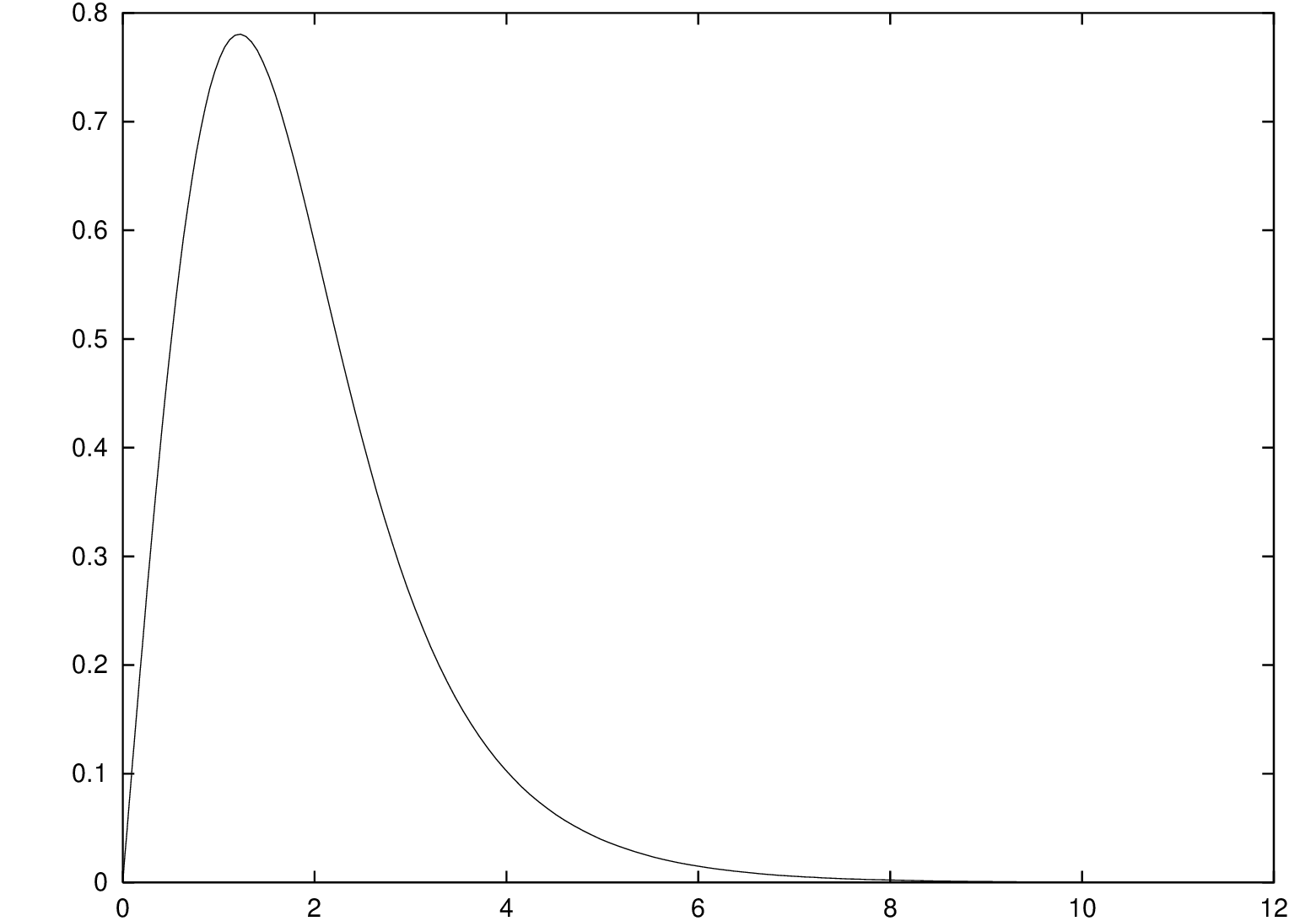}}
\end{figure}
\clearpage

\begin{figure}
\centerline{\epsfysize=10cm\epsffile{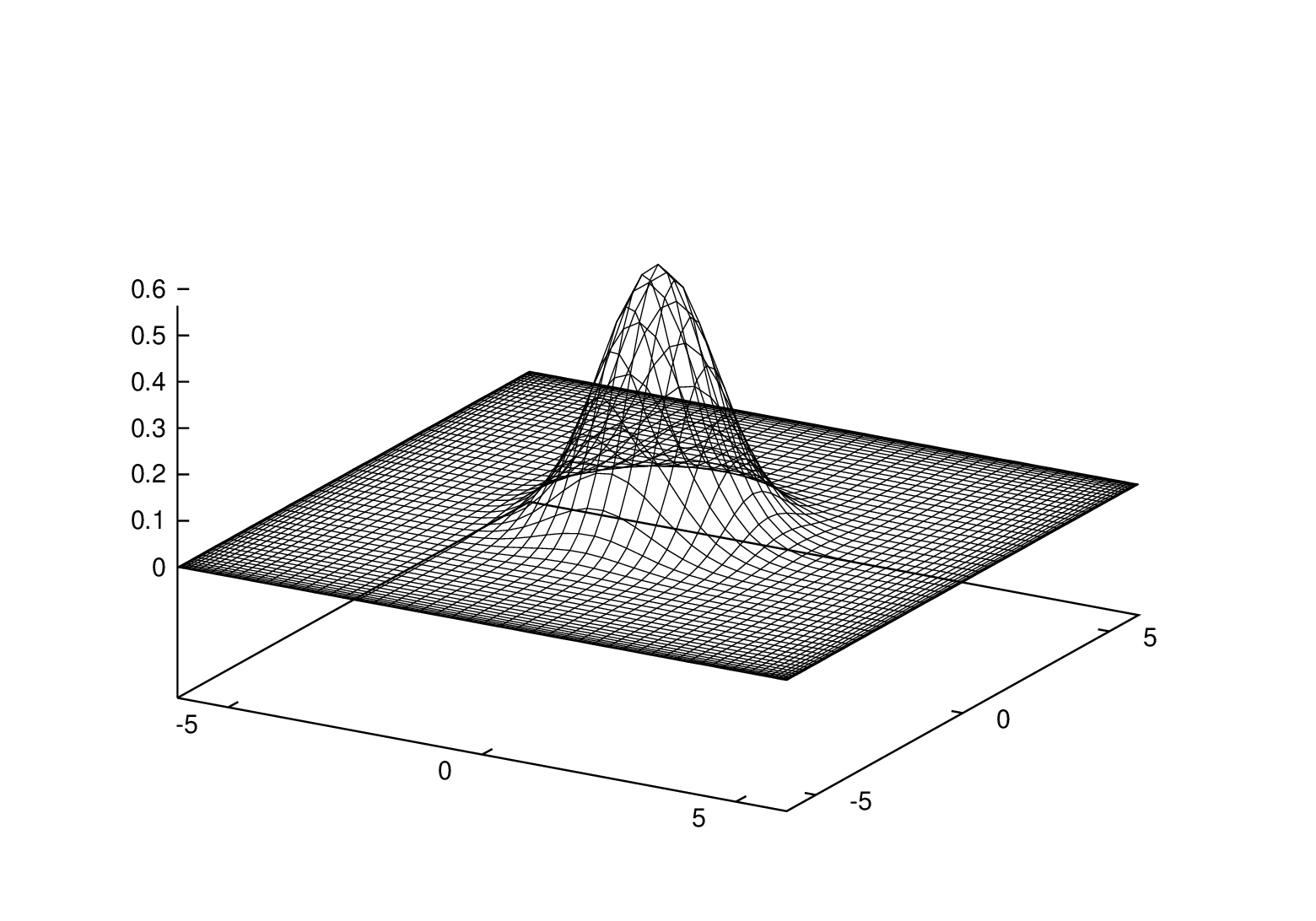}}
\end{figure}
\clearpage

\begin{figure}
\centerline{\epsfysize=10cm\epsffile{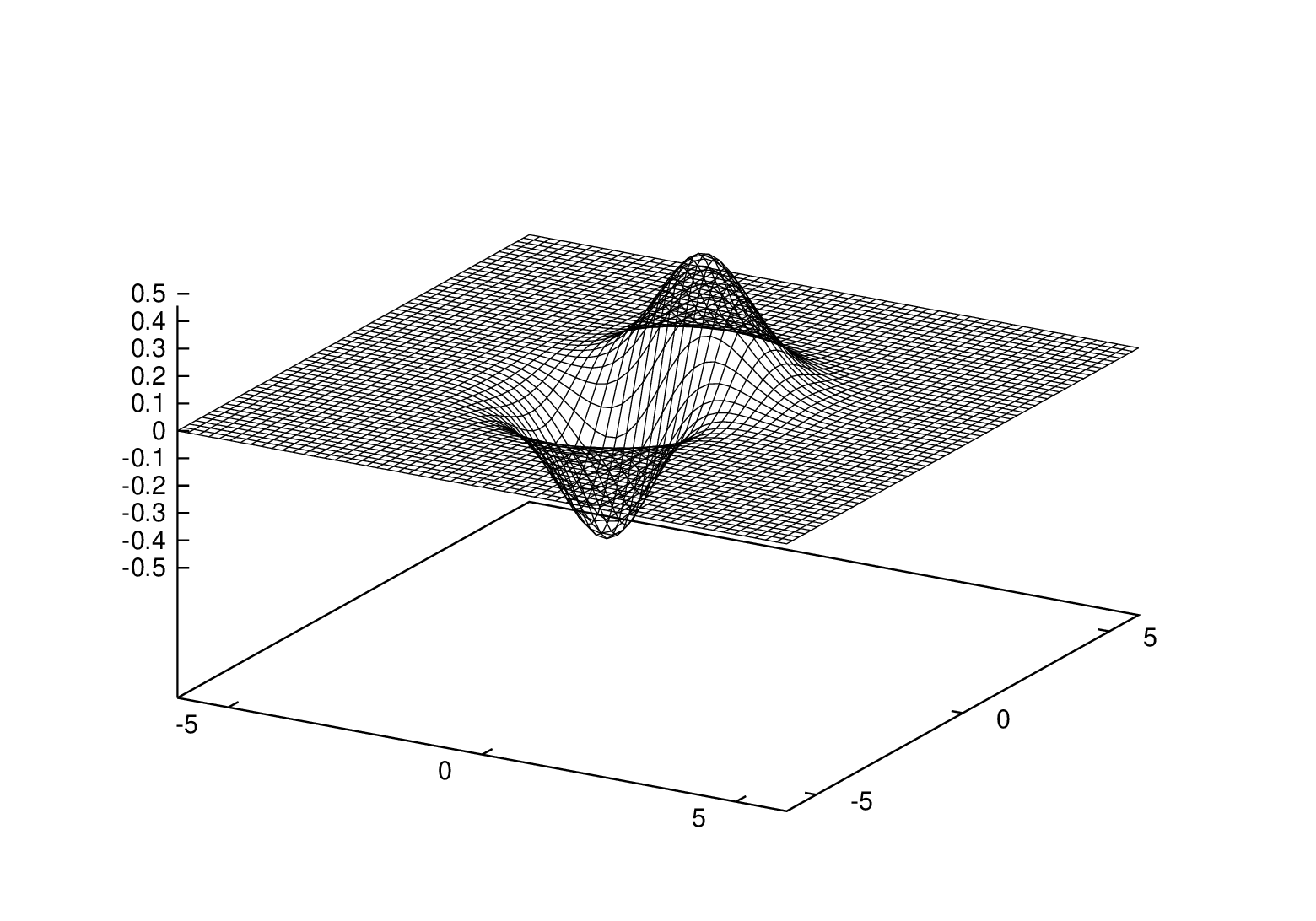}}
\end{figure}
\clearpage

\begin{figure}
\centerline{\epsfysize=10cm\epsffile{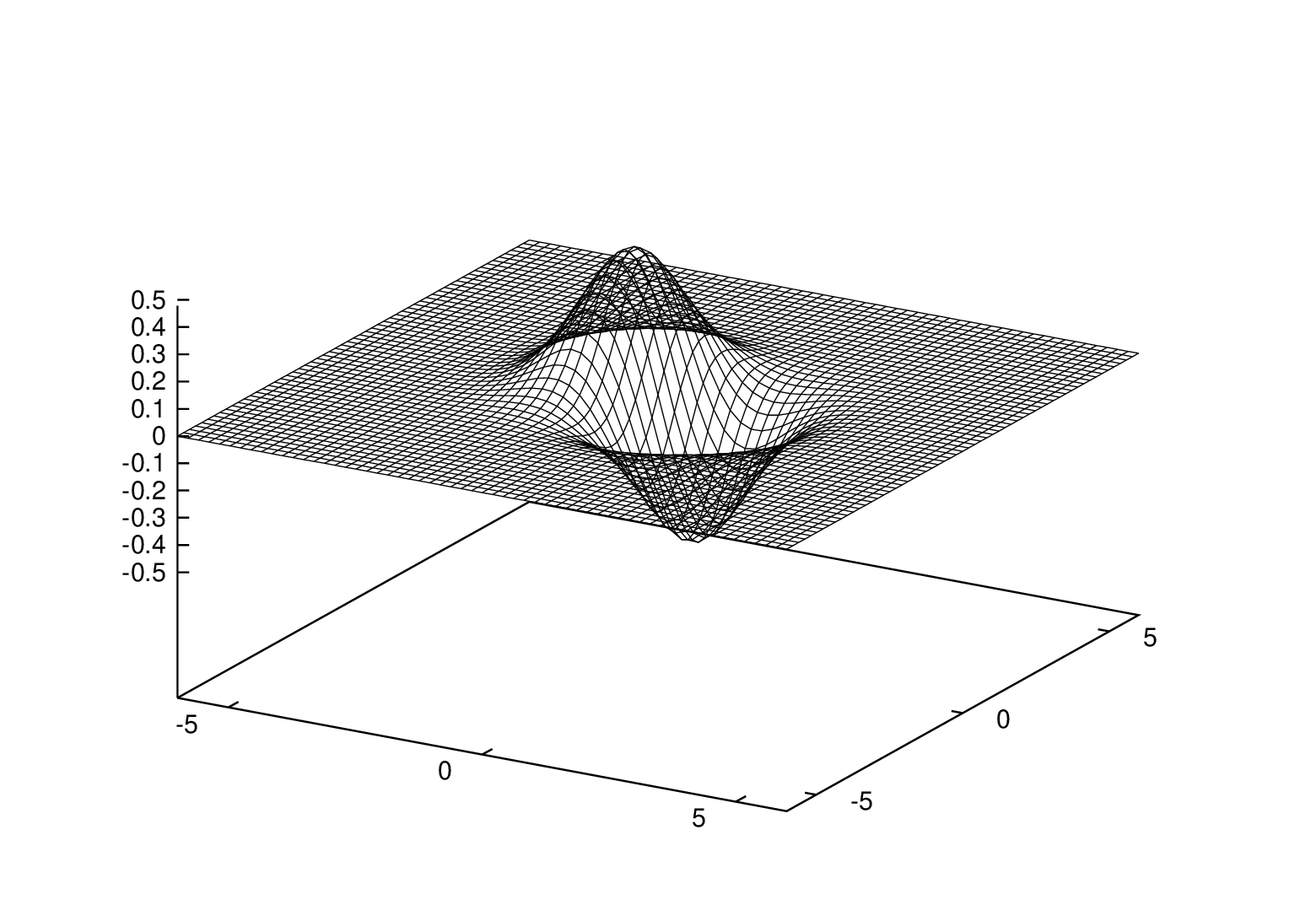}}
\end{figure}
\clearpage

\begin{figure}
\centerline{\epsfysize=10cm\epsffile{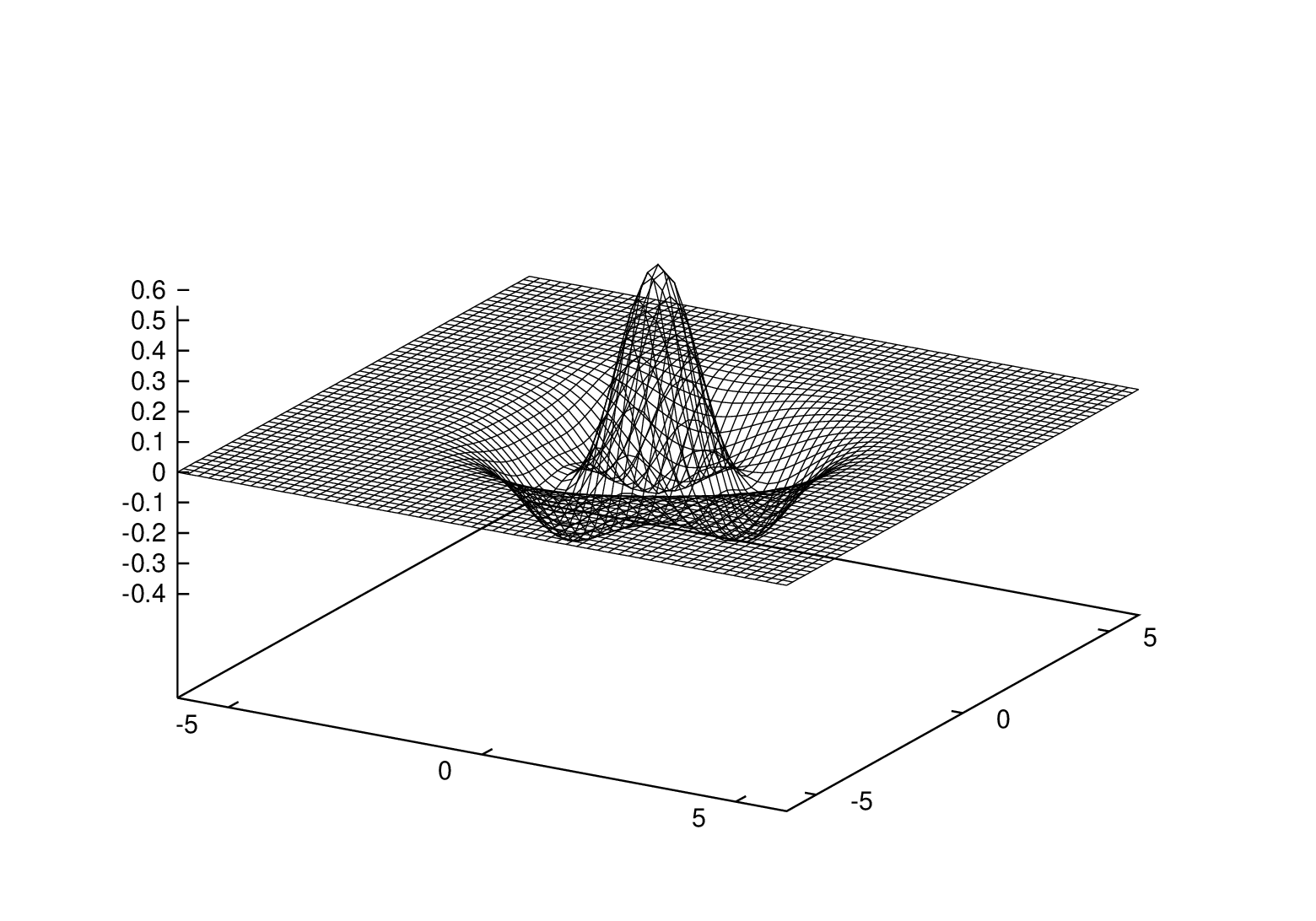}}
\end{figure}
\clearpage

\begin{figure}
\centerline{\epsfysize=10cm\epsffile{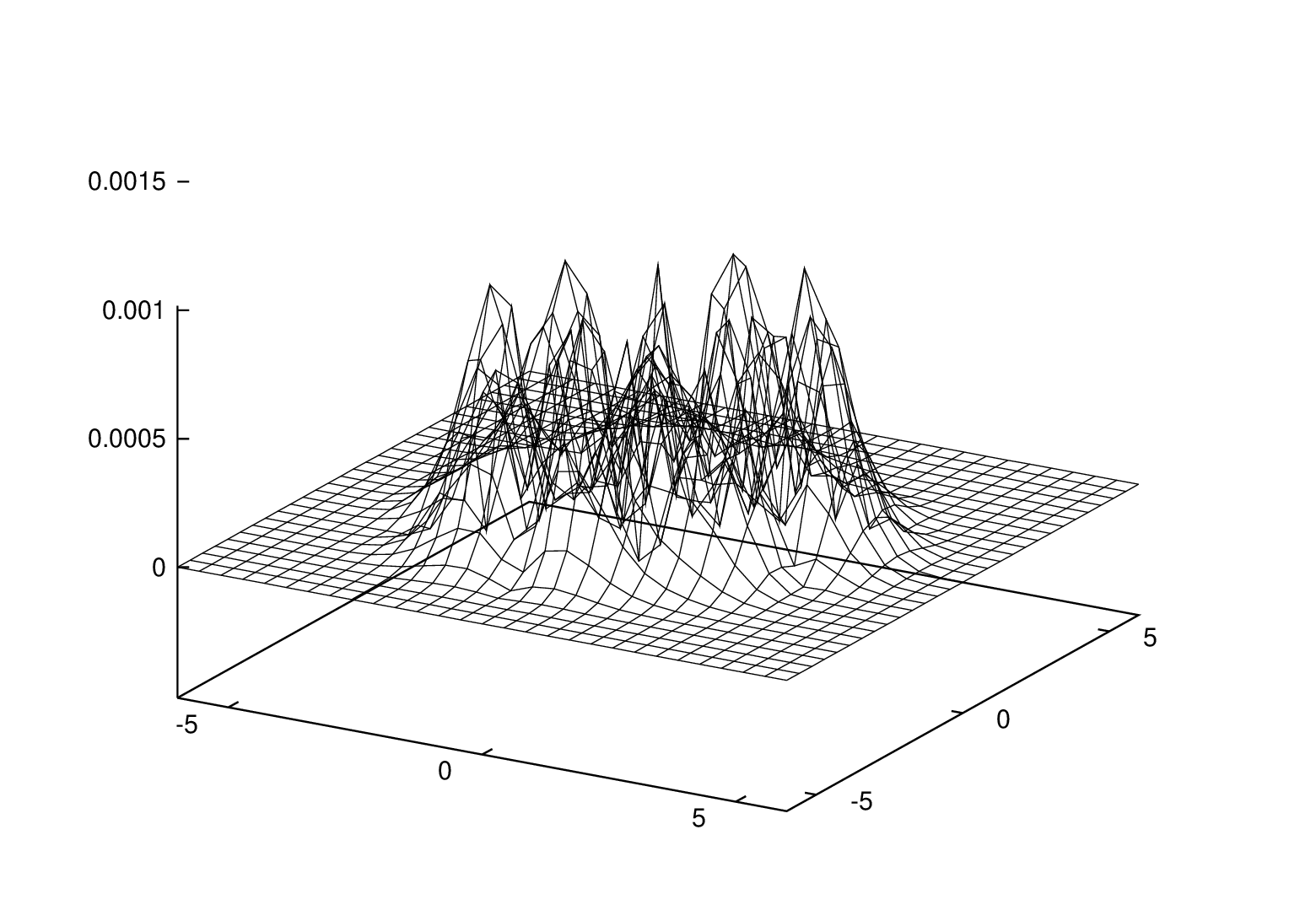}}
\end{figure}
\clearpage

\begin{figure}
\centerline{\epsfysize=10cm\epsffile{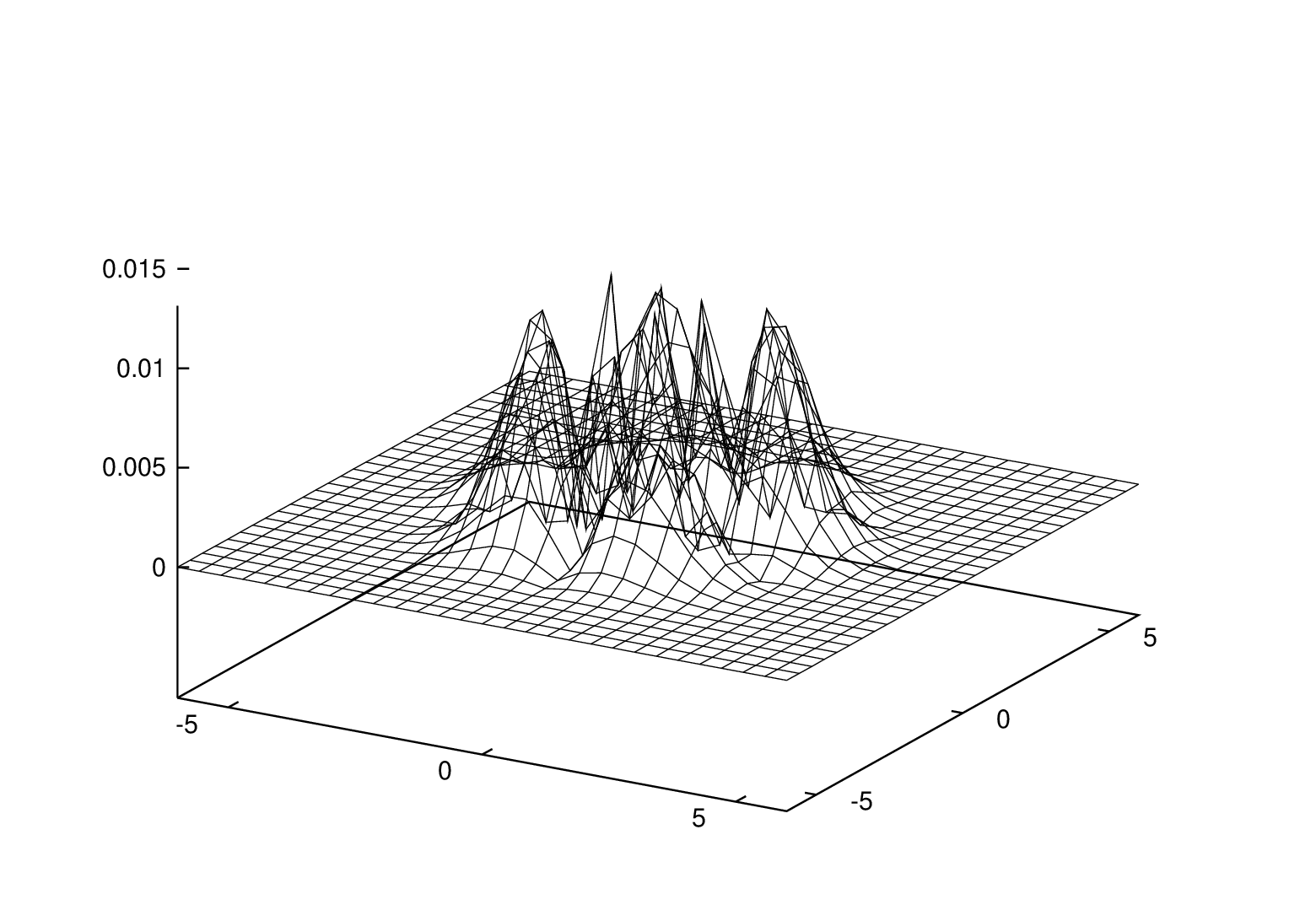}}
\end{figure}
\clearpage

\begin{figure}
\centerline{\epsfysize=10cm\epsffile{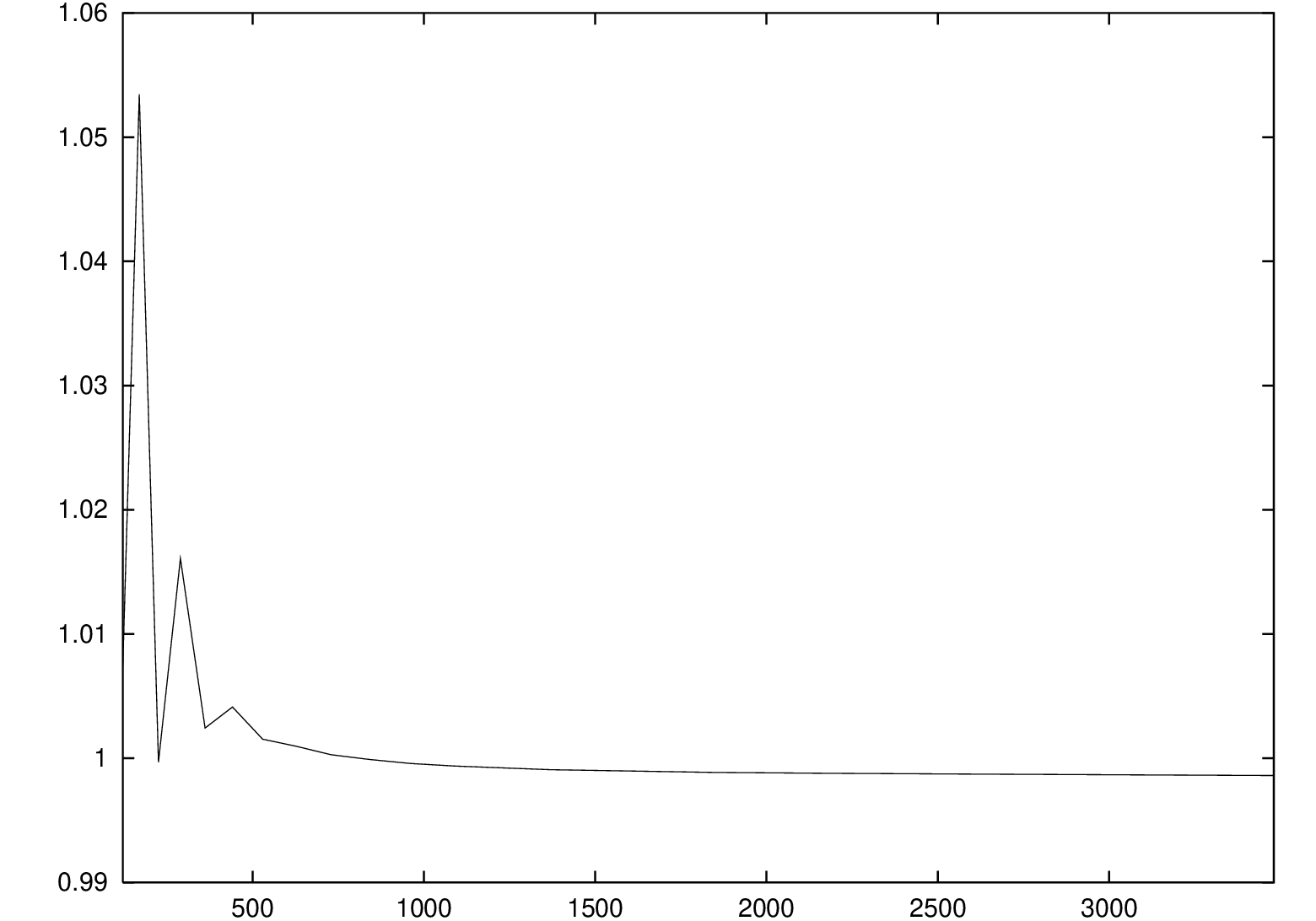}}
\end{figure}
\clearpage

\begin{figure}
\centerline{\epsfysize=10cm\epsffile{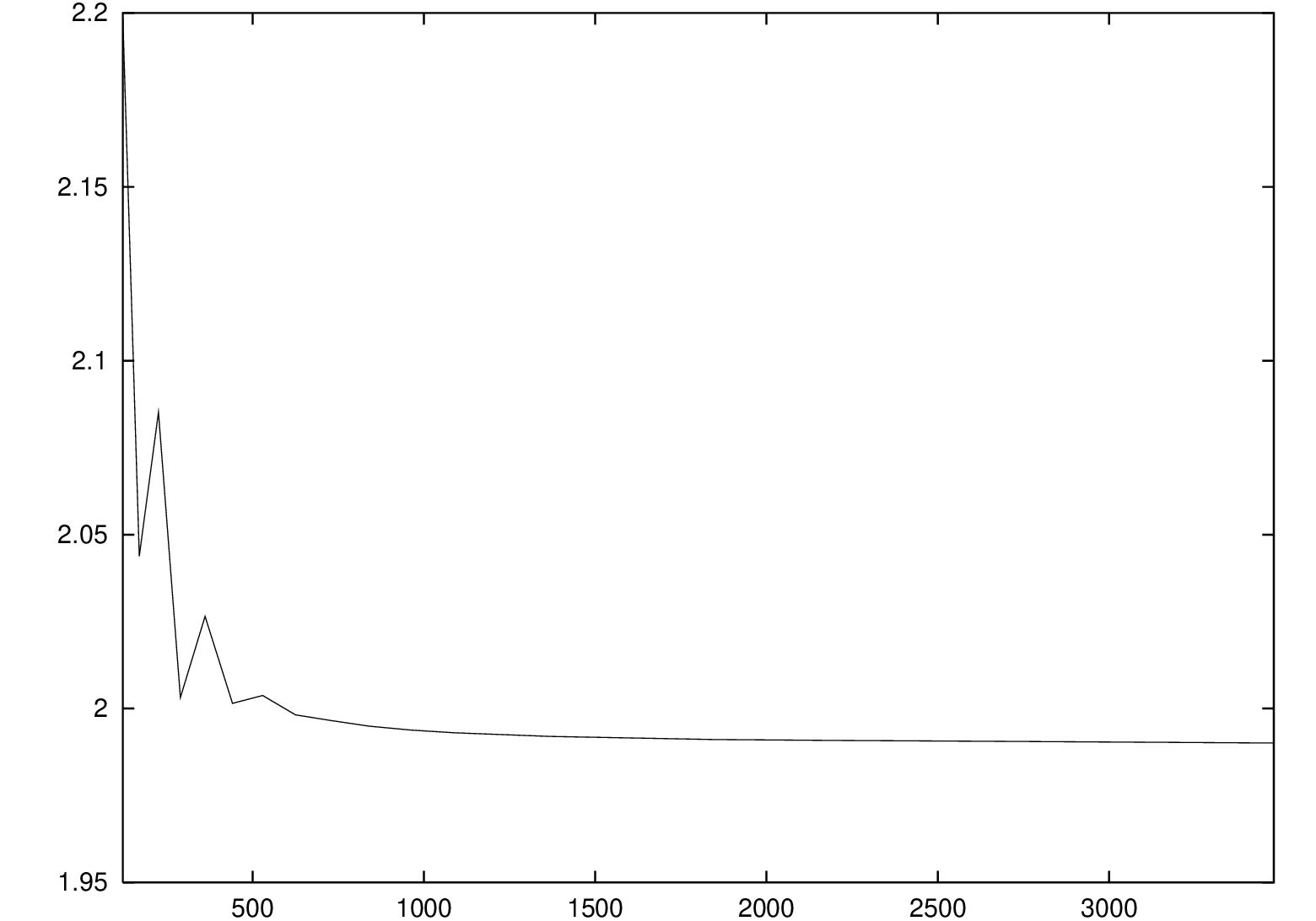}}
\end{figure}

\end{document}